# Probabilistic Interval Analysis for the Analytic Prediction of the Pattern Tolerance Distribution in Linear Phased Arrays with Random Excitation Errors

Paolo Rocca, *Senior Member, IEEE*, Nicola Anselmi, *Member, IEEE*, Arianna Benoni,
and Andrea Massa, *Fellow, IEEE*

*Abstract*—A statistical approach based on the Interval Analysis (*IA*) is proposed for the analysis of the effects, on the radiation patterns radiated by phased arrays, of random errors and tolerances in the amplitudes and phases of the array-elements excitations. Starting from the efficient, reliable, and inclusive computation of the bounds of the complex-valued interval array pattern function by means of *IA*, an analytic method is presented to yield closed-form expressions for the probability of occurrence of a user-chosen value of the power pattern or of its features within the corresponding *IA*-derived bounds. A set of numerical examples is reported and discussed to assess the reliability of the proposed probabilistic interval analysis (*PIA*) method with the results from Monte Carlo simulations as well as to point out its effectiveness and potentialities/advantages/efficiency in real applications of great industrial interest.

*Index Terms*—Phased Array Antenna, Linear Arrays, Amplitude and Phase Excitation Errors, Tolerance Analysis, Probabilistic Interval Analysis.

## I. INTRODUCTION

NOWADAYS, the assessment and - even more - the prediction of the impact on the antenna array performance of the errors and/or the tolerances caused by the fabrication process [1][2], the material defects, and the operation/working conditions (e.g., electromagnetic couplings [3], mechanical strains [4], and thermal drifts [5]) is of great importance. Indeed, phased arrays (*PA*s) are and they will be increasingly used for key commercial and industrial applications [6] (e.g., 5G and next-generation mobile communications, autonomous driving, and industry 4.0) as an effective technological solution to guarantee high-quality and reliable data links thanks to the synthesis of complex pattern features (e.g., directive mainlobes, low sidelobes, etc.) through a precise control of the array element excitations. Because of the ever growing number of systems and scenarios that exploit *PA*-based architectures, a number of uncertainties, also totally new, will certainly occur with unknowns effects. In order to predict the differences in the behavior of the real/actual implementation from its numerical model, several tolerance analysis techniques, based on either statistical or analytic/interval methods, have been proposed since the origin of *PA*s to analyze the effects of the deviations from the nominal values of the excitation coefficients. In the state-of-the-art literature, statistical approaches [7]-[12] provide simple closed-form expressions for the mean value of the power pattern and its main features [e.g., mainlobe peak, sidelobe level (*SLL*), and pattern nulls] by exploiting the central limit theorem (*CLT*) and small error approximations. Despite the wide application, due to their simplicity and effectiveness when the underlying hypotheses hold true (e.g., the case of large enough arrays, such that the *CLT* is valid, and small errors[1]), statistical techniques have not always been reliable when dealing with small/medium arrays [13]. In order to numerically estimate the probability distribution function of the power pattern and of the array features, Monte Carlo simulations [14] have been exploited, as well. Although such a methodological approach does not recur to any approximation and it allows the testing of realistic arrays modeled with very accurate computer-aided simulation tools, the intrinsic need to run a statistically meaningful number of simulations turns out to be prohibitive because of the rapidly growing computational burden due to the huge set of admissible error configurations also in case of small arrays. As for the computation of the

---

Manuscript received on January XX, 2020; revised on May XX, 2020

This work benefited from the networking activities carried out within the Project "CYBER-PHYSICAL ELECTROMAGNETIC VISION: Context-Aware Electromagnetic Sensing and Smart Reaction (EMvisioning)" (Grant no. 2017HZJXSZ) funded by the Italian Ministry of Education, University, and Research within the Program PRIN2017 (CUP: E64I19002530001) and the Project "WATERTECH - Smart Community per lo Sviluppo e l'Applicazione di Tecnologie di Monitoraggio Innovative per le Reti di Distribuzione Idrica negli usi idropotabili ed agricoli" (Grant no. SCN_00489) funded by the Italian Ministry of Education, University, and Research within the Program "Smart cities and communities and Social Innovation" (CUP: E44G14000060008).

P. Rocca, N. Anselmi, A. Benoni, and A. Massa are with the ELEDIA@UniTN (DISI - University of Trento), Via Sommarive 9, 38123 Trento - Italy (e-mail: {paolo.rocca, nicola.anselmi.1, arianna.benoni, andrea.massa}@unitn.it).

P. Rocca, is also with the ELEDIA Research Center (ELEDIA@XIDIAN - Xidian University),P.O. Box 191, No.2 South Tabai Road, 710071 Xi'an, Shaanxi Province - China (e-mail: paolo.rocca@xidian.edu.cn).

A. Massa is also with the ELEDIA Research Center (ELEDIA@L2S - UMR 8506), 3 rue Joliot Curie, 91192 Gif-sur-Yvette - France (e-mail: andrea.massa@l2s.centralesupelec.fr).

A. Massa is also with the ELEDIA Research Center (ELEDIA@UESTC - University of Electronic Science and Technology of China), School of Electronic Engineering, 611731 Chengdu - China (e-mail: andrea.massa@uestc.edu.cn).

A. Massa is also with the ELEDIA Research Center (ELEDIA@TSINGHUA - Tsinghua University), 30 Shuangqing Rd, 100084 Haidian, Beijing - China (e-mail: andrea.massa@tsinghua.edu.cn).

---

[1]For example the small phase approximation, $e^{jx} \simeq 1 + x$, can be used for relatively small ($x \ll 1$) deviations in the phase of the array excitations or in the position of the array elements [12].

probability of exceeding a user-chosen *SLL* value, improved statistical models, devoted to describe the distribution of the magnitude of the array amplitude pattern, have been proposed [15], as well.

Recently, analytic techniques based on the arithmetic of intervals and on the theory of the Interval Analysis [16][17] have been introduced to efficiently compute the tolerance bounds of the radiated power pattern and of the related features of an antenna affected by errors and uncertainties laying within known intervals of values. More specifically, the key features of such *IA*-derived bounds are that they are finite and inclusive, since all the admissible realizations of the power pattern of the actual antenna lie within the *IA* interval, thanks to the *IA Inclusion Property* [16][17]. Moreover, it is worth pointing out that the *IA* bounds are exact since yielded by extending the crisp power pattern function to the corresponding interval function without any approximation and without recurring to an exhaustive (ideally infinite) run of Monte Carlo simulations. Indeed, the only quantities involved in the *IA*-based maths are the endpoints of the intervals of the input variables (i.e., the antenna descriptors such as the excitations for the antenna arrays) affected by uncertainties. Thanks to these advantages, *IA*-based methods have been profitably applied to different antenna systems and devices such as *PA*s [18]-[23], reflector antennas [24][25], reflectarrays, [26] and antenna materials [27] or radomes [28][29]. Moreover, *IA*-based tolerance analysis techniques have been exploited, jointly with optimization algorithms, for the robust synthesis of antennas [30][31][32] in order to obtain antenna designs resilient to uncertainties within the considered error intervals, without the need of correcting their effects through suitable compensation methods [33]-[35]. Thanks to its effectiveness, *IA* has been also recently applied with success to thermal structures [36], composite laminates [37], and the predictions of the structures behaviour [38].

By focusing on the tolerance analysis of *PA* through *IA* methods, a Cartesian implementation has been firstly introduced to separately evaluate the effects on the array performance of amplitude [18][19] and phase [13] excitation errors. Analytic closed-form expressions for the infimum and the supremum of the real and imaginary parts of the interval array factor function have been derived in terms of only the endpoints of the excitations errors. Successive extensions have dealt with calibration errors and mutual coupling effects through the circular interval arithmetic and the circular *IA* in which the intervals are represented as circles in the complex plane centered in the nominal value and characterized by an interval radius proportional to the error tolerance [39]. To mitigate the dependency problem[2], a Taylor expansion has been used to estimate the sensitivity of the array pattern performance on small errors in the excitation amplitudes [21], while a matrix-based *IA* method has been applied to *PA*s with errors on the amplitude coefficients [22][23]. The presence of both amplitude and phase deviations from the nominal array weights has been addressed in [20] by means of a suitable integration of the *IA* theory with a Minkowski-Sum procedure (*IA-MS*). Such a technique has proved to be able to mitigate the overestimation of the power pattern bounds, of both the Cartesian and the circular *IA* approaches, due to the "wrapping effect" in the interval representation of the complex intervals/phasors for the computation of the array radiation pattern.

Despite the significant advances of *IA*-based methods over other state-of-the-art competitive alternative for sensitivity analysis, the key limitation of such analytic/inclusive approaches is that, to date and to the best of the authors' knowledge, no information on the probability of occurrence of a value within the interval bounds is available, but only the upper and the lower limits of the power pattern function are available. This work is aimed at overcoming such a drawback by presenting a Probabilistic Interval Analysis (*PIA*) method that provides, through analytic rules, the probability distribution of the power pattern values and of the array performance indexes within the corresponding *IA*-computed bounds. More specifically, starting from the accurate description of the complex intervals with the *IA-MS* method, the information coded into the representation of the *PA* radiation pattern in the complex plane, as a function of the angular direction and of the nominal array excitations together with their tolerance intervals, is mapped into a probability distribution function (*PDF*). The main methodological novelties of this research study over the tolerance methods available in the literature comprise (*i*) the introduction of an innovative *IA*-based strategy to give a probabilistic distribution of the occurrence of the power pattern values within inclusive, yet finite, bounds for the tolerance analysis of *PA*s characterized by arbitrary, but bounded, amplitude and phase excitation errors; (*ii*) the complete development of a customized procedure, which is based on analytic relationships, for the computation of the power pattern *PDF* by extracting the probabilistic information embedded into the complex envelope of the interval function of the array radiation pattern.

The rest of the paper is organized as follows. The *PIA* method is formulated in Sect. II by also detailing the numerical procedure for computing the *PDF* of the power pattern within the *IA-MS* bounds (Sect. II.A). Section III is devoted to the numerical analysis and the assessment of the proposed approach also in comparison with the Monte Carlo method. Finally, some conclusions and final remarks are drawn (Sect. IV).

## II. MATHEMATICAL FORMULATION

Let us consider a linear phased array of $N$ isotropic elements uniformly-spaced by $d$ along the $x$-axis and controlled in amplitude and phase through a set of $N$ amplifiers and $N$ phase shifters, respectively. Let the nominal (complex) excitation of the $n$-th ($n = 1, ..., N$) array element be

$$W_n = A_n e^{jB_n}, \quad (1)$$

$A_n$ and $B_n$ being its nominal (i.e., error/tolerance-free) amplitude and phase coefficient, respectively. The far-field pattern

---
[2]The dependency problem arises when an interval variable is present more than once into an expression. Without suitable mathematical manipulations, this causes an overestimation of the bounds of the result of the expression at hand since each occurrence of the same interval variable is considered as independent (i.e., a different variable).

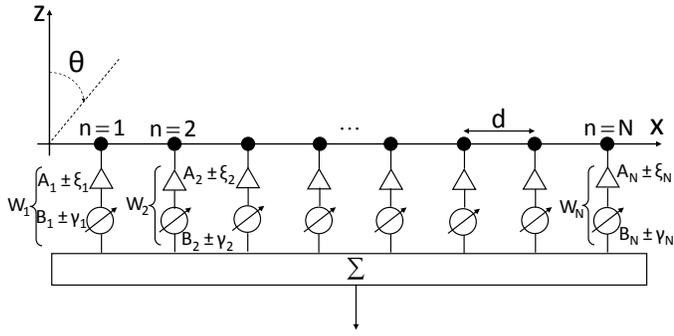

Figure 1. Sketch of a linear *PA* with $N$ elements uniformly-spaced by $d$ along the $x$-axis.

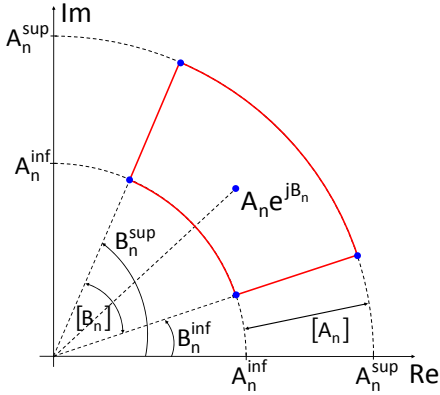

Figure 2. Representation in the complex plane of the $n$-th ($n = 1, ..., N$) complex-valued excitation interval $[\widetilde{W}_n]$ ($[W_n] \triangleq [A_n] e^{j[B_n]}$; $[W_n] \triangleq \left[W_n^{inf}, W_n^{sup}\right]$) as a function of the real-valued amplitude $[A_n]$ ($[A_n] \triangleq \left[A_n^{inf}, A_n^{sup}\right]$) and phase $[B_n]$ ($[B_n] \triangleq \left[B_n^{inf}, B_n^{sup}\right]$) intervals.

of the array is mathematically described by the array factor (*AF*)

$$AF(u) = \sum_{n=1}^{N} A_n e^{jB_n} e^{j\frac{2\pi}{\lambda}d(n-1)u} \quad (2)$$

where $j = \sqrt{-1}$ is the complex variable, $\lambda$ is the wavelength, and $u = \sin(\theta)$ is the angular direction (Fig. 1).

Let us suppose that the amplifiers and the phase shifters, which determine/control the amplitude and the phase excitation weights in the beamforming network, are affected by known or measurable tolerances (i.e., deviations from the nominal value) such that the actual $n$-th ($n = 1, ..., N$) array coefficients, $\widetilde{A}_n$ and $\widetilde{B}_n$, belong to the interval [16][17] $[A_n]$

$$[A_n] \triangleq [A_n^{inf}, A_n^{sup}] = [A_n - \xi_n^{inf}, A_n + \xi_n^{sup}] \quad (3)$$

and $[B_n]$

$$[B_n] \triangleq [B_n^{inf}, B_n^{sup}] = [B_n - \gamma_n^{inf}, B_n + \gamma_n^{sup}], \quad (4)$$

respectively ($\widetilde{A}_n \in [A_n]$, $\widetilde{B}_n \in [B_n]$), $\xi_n^{inf/sup}$ and $\gamma_n^{inf/sup}$ being the maximum deviations from the corresponding nominal values. Accordingly, the $n$-th ($n = 1, ..., N$) actual complex excitation, $\widetilde{W}_n$, lies within the complex interval $[W_n]$ given by

$$[W_n] \triangleq [A_n] e^{j[B_n]} \quad (5)$$

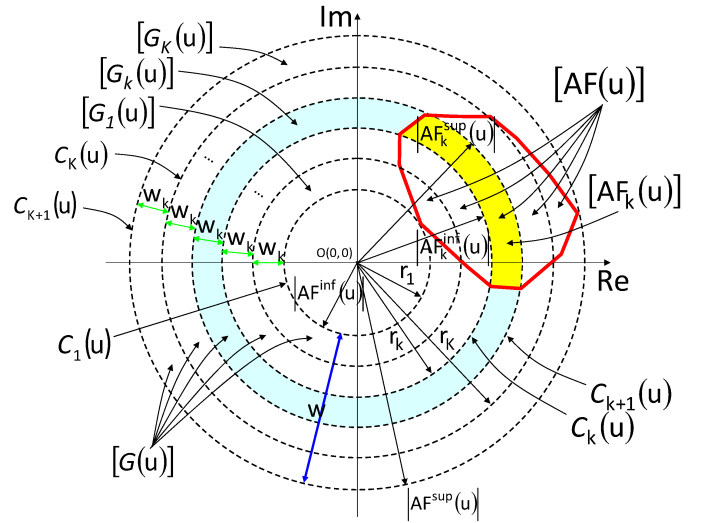

Figure 3. Representation in the complex plane of the complex-valued interval functions $[G(u)]$ ($[G(u)] \triangleq [G^{inf}(u), G^{sup}(u)]$) and $[AF(u)]$ (i.e., the interval array factor, $[AF(u)] \triangleq [AF^{inf}(u), AF^{sup}(u)]$) along with their $K$ partitions, $[G_k(u)]$ and $[AF_k(u)]$ ($k = 1, ..., K$), so that $[G(u)] = \cup_{k=1}^{K} [G_k(u)]$, $[AF(u)] = \cup_{k=1}^{K} [AF_k(u)]$, and $[AF_k(u)] = [AF(u)] \cap [G_k(u)]$, while $G_k^{inf}(u)$ and $G_k^{sup}(u)$ coincide with the circles $C_k(u)$ and $C_{k+1}(u)$ centered in the origin and having radius $r_k = |AF^{inf}(u)| + (k-1)w_k$ ($\to G_k^{inf}(u) = r_k$) and $r_{k+1} = |AF^{inf}(u)| + kw_k$ ($\to G_k^{sup}(u) = r_{k+1}$), respectively, being $r_1 = |AF^{inf}(u)|$ and $r_{K+1} = |AF^{sup}(u)|$.

as shown in Fig. 2.

By substituting (5) into (2), the *AF* turns out to be a complex interval function $[AF(u)]$ defined as

$$[AF(u)] \triangleq \sum_{n=1}^{N} [W_n] e^{j\frac{2\pi}{\lambda}d(n-1)u} \quad (6)$$

and characterized by an infimum/lower-bound, $AF^{inf}(u)$, and a supremum/upper-bound, $AF^{sup}(u)$, crisp functions of the angular variable $u$ (i.e., $[AF(u)] \triangleq [AF^{inf}(u), AF^{sup}(u)]$).

According to the *IA-MS* method [20], the complex interval $[AF(u)]$ is represented through a polygon (Fig. 3), which is the smallest convex hull of the interval phasor, whose vertices are computed according to the Minkowski Sum (*MS*) strategy [40]. Afterwards, the lower and the upper bounds of the real-valued power pattern interval function, $[P(u)] \triangleq |[AF(u)]|^2$, are derived by firstly determining the bounds of the *AF* module, $|[AF(u)]|$ ($|[AF(u)]| \triangleq [|AF^{inf}(u)|, |AF^{sup}(u)|]$), as the minimum, $|AF^{inf}(u)|$, and the maximum, $|AF^{sup}(u)|$, distance of the corresponding complex interval $[AF(u)]$ from the origin of the complex plane (Fig. 3). Finally, the bounds of $[P(u)]$ are easily yielded by setting $P^{inf}(u) = |AF^{inf}(u)|^2$ and $P^{sup}(u) = |AF^{sup}(u)|^2$.

Although the *IA-MS* significantly improves the accuracy in predicting the bounds of $[P(u)]$ with respect to [13][18][39], it is worth pointing out that the interval polygon $[AF(u)]$ ($u \in [-1, 1]$), resulting from the complex-valued interval operations, contains more information than the one needed for computing $[P(u)]$. For instance, the two complex intervals, $[AF(u_1)]$ and $[AF(u_2)]$, with the same $AF$ module interval $|[AF(u)]|$ [i.e., $|[AF(u_1)]| = |[AF(u_2)]| \rightarrow |AF^{inf}(u_1)| = |AF^{inf}(u_2)|$ and $|AF^{sup}(u_1)| = |AF^{sup}(u_2)|$ - Fig. 4($a$)] turn out to have the same power pattern interval [i.e., $[P(u_1)] = [P(u_2)]$ - Fig. 4($b$)], but they identify two different regions/shapes in the complex plane [i.e., $[AF(u_1)] \neq [AF(u_2)]$ - Fig. 4($a$)]. The probability that the power pattern of a given array with specified $AF$, $AF(u)$, assumes a value $\hat{P}(u)$ within $[P(u)]$ ($p^P(u) \triangleq \Pr\{\hat{P}(u) \in [P(u)]\}$) is expected to be different for each complex interval $[AF(u_1)]$ and $[AF(u_2)]$ [i.e., $p^P(u_1) \neq p^P(u_2)$ - Fig. 4($e$)] since it is related to the membership, in the complex plane, of the corresponding $AF$ sample $\hat{AF}(u)$ to $[AF(u)]$ [e.g., Figs. 4($c$)-4($d$)] being $\hat{P}(u) \triangleq |\hat{AF}(u)|^2$.

In the following, the information coming from the knowledge of $[AF(u)]$, but not used for computing $[P(u)]$, is exploited to analytically predict the power pattern tolerance distribution.

### A. Probabilistic Interval Analysis (PIA) Method

Starting from the complex-valued interval array factor function (6) obtained with the *IA-MS* approach [20], the following procedure is applied to compute the *PDF* associated to the occurrence of a power pattern value $\hat{P}(u)$, along the angular direction $u$ ($u \in [-1,1]$), within the interval bounds $P^{inf}(u)$ and $P^{sup}(u)$. More specifically, the following steps are performed:

- **Step 1 - Identification of the Probability Regions**. The interval bounds $|AF^{inf}(u)|$ and $|AF^{sup}(u)|$ define in the complex plane an interval ring (or a circle if $|AF^{inf}(u)| = 0$) $[G(u)]$ of width $w = |AF^{sup}(u)| - |AF^{inf}(u)|$ to which the interval phasor $[AF(u)]$ belongs to (Fig. 3). Such a ring is then partitioned into $K$ uniform annular regions, $\{[G_k(u)]; \ k = 1, ..., K\}$ so that $[G(u)] = \cup_{k=1}^{K} [G_k(u)]$ and the width of each $k$-th ($k = 1, ..., K$) sub-ring is equal to $w_k = \frac{w}{K}$ (Fig. 3). Moreover, the lower, $G_k^{inf}(u)$, and the upper, $G_k^{sup}(u)$, bounds of the $k$-th ($k = 1, ..., K$) annular region $[G_k(u)]$ are concentric circles $C_k(u)$ and $C_{k+1}(u)$ centered in the origin of the complex plane with radius

$$r_k = |AF^{inf}(u)| + (k-1)w_k \quad (7)$$

and

$$r_{k+1} = |AF^{inf}(u)| + kw_k, \quad (8)$$

respectively, subject to $r_1 = |AF^{inf}(u)|$ and $r_{K+1} = |AF^{sup}(u)|$. Therefore, the region that identifies the interval $AF$ in the complex plane turns out to be subdivided into $K$ planar sectors, $\{[AF_k(u)]; \ k = 1, ..., K\}$, that by definition fit the condition $[AF(u)] = \cup_{k=1}^{K} [AF_k(u)]$, the $k$-th ($k = 1, ..., K$) sub-domain being the intersection of the interval $AF$, $[AF(u)]$, with the $k$-th annular region $[G_k(u)]$

$$[AF_k(u)] = [AF(u)] \cap [G_k(u)] \quad (9)$$

where $\cap$ stands for the intersection operator;

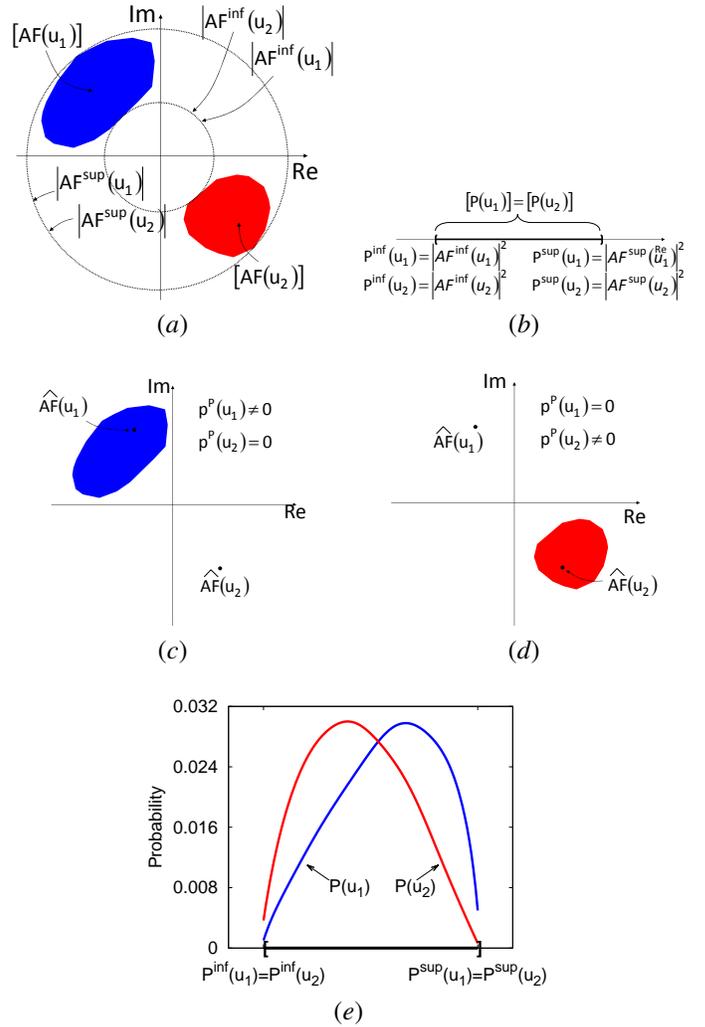

Figure 4. Illustrative example of the relationships between ($a$) the array factor interval $[AF(u)]$, ($b$) the power pattern interval $[P(u)]$, and ($c$)-($e$) the probability of power pattern occurrence, $p^P(u)$.

- **Step 2 - Computation of the PDF of $|AF(u)|$**. The *PDF* of $|AF(u)|$, $p_k^{|AF|}(u)$, is analytically determined as a step-wise function of the $K$ strips in the complex plane. Since it defines the probability that the amplitude of the array factor assumes a value $|\hat{AF}(u)|$ within the $k$-th ($k = 1, ..., K$) region $[|AF_k(u)|]$

$$p_k^{|AF|}(u) \triangleq \Pr\{|\hat{AF}(u)| \in [|AF_k(u)|]\}, \quad (10)$$

it is related to the ratio between the area of the $k$-th ($k = 1, ..., K$) annular sector $[AF_k(u)]$ and that of the whole region $[AF(u)]$

$$p_k^{|AF|}(u) = \frac{\mathcal{A}\{[AF_k(u)]\}}{\mathcal{A}\{[AF(u)]\}} \quad (11)$$

where $\mathcal{A}(\cdot)$ stands for the operator returning the area of the argument;

- **Step 3 - Computation of the Occurrence Probability $P(u)$**. Since each sample $|\hat{AF}(u)|$ ($|\hat{AF}(u)| \in [|AF_h(u)|]$; $h \in [1, K]$) maps into a corresponding power pattern one $\hat{P}(u)$ ($\hat{P}(u) \in [|P_h(u)|]$; $h \in$

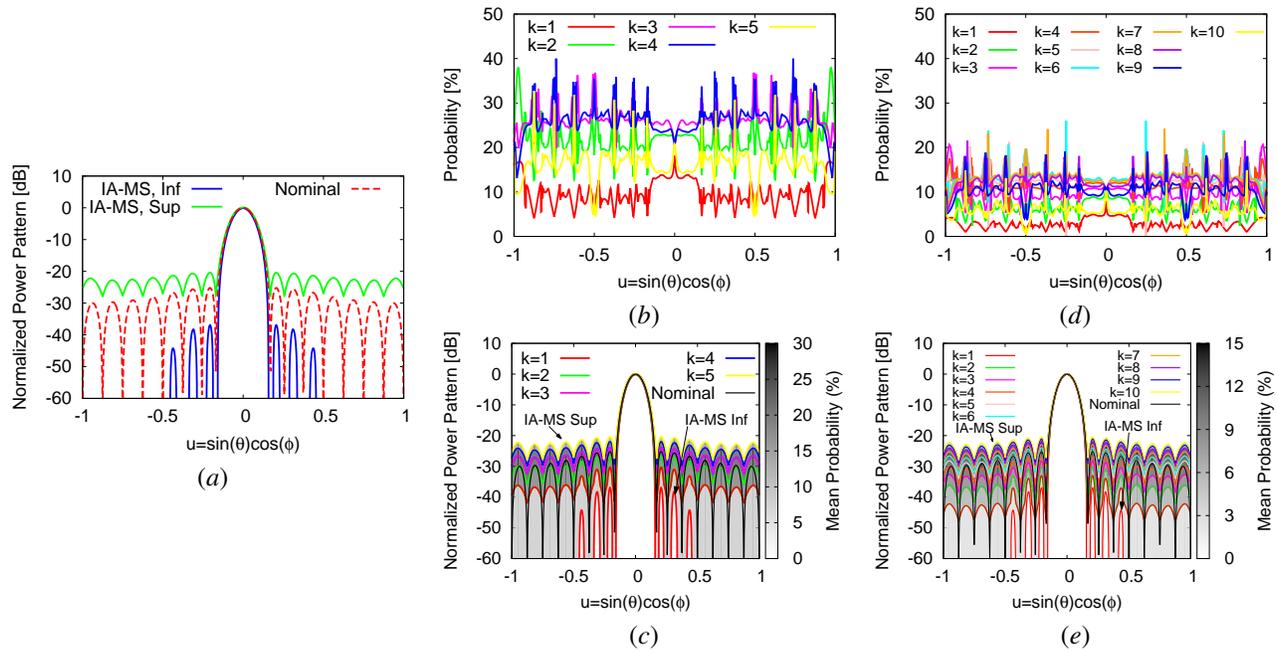

Figure 5. *Test Case 1* ($N = 16$, $d = \lambda/2$, $\xi = 1$ %, $\gamma = 3$ [deg]; Taylor pattern: $SLL = -25$ [dB], $\bar{n} = 3$) - Plot of (a) the *IA-MS* computed bounds of the power pattern interval function, $[P(u)]$, together with the nominal power pattern function, $P(u)$, (b)(d) the probability functions $\{p_k^P(u); k = 1, ..., K\}$, and (c)(e) the $K$ power pattern probability regions $\{[P_k(u)]; k = 1, ..., K\}$ with the grey-level representation of the mean probability values $\{\bar{p}_k; k = 1, ..., K\}$ when (b)(c) $K = 5$ and (d)(e) $K = 10$.

$[1, K])$ and $\hat{P}(u) \triangleq \left|\hat{AF}(u)\right|^2$ being $[P_k(u)] \triangleq \left[P_k^{inf}(u), P_k^{sup}(u)\right]$ $(k = 1, ..., K)$ where $P_k^{inf}(u) = \left|AF_k^{inf}(u)\right|^2$ and $P_k^{sup}(u) = \left|AF_k^{sup}(u)\right|^2$ $(k = 1, ..., K)$, it turns out that the probability that the power pattern assumes a value $\hat{P}(u)$ within the $k$-th $(k = 1, ..., K)$ region $[P_k(u)]$

$$p_k^P(u) \triangleq \Pr\left\{\hat{P}(u) \in [P_k(u)]\right\} \quad (12)$$

is equal to (10), $p_k^P(u) = p_k^{|AF|}(u)$, thus its value can be computed as in (11) and according to the numerical procedure detailed in *Appendix I*. Moreover, the mean value of the probability of occurrence of the $k$-th ($k = 1, ..., K$) power pattern region $[P_k(u)]$, $\bar{p}_k$, can be yielded as follows

$$\bar{p}_k = \frac{1}{2}\int_{u=-1}^{+1} p_k^P(u)\,du \quad (13)$$

physically indicating the mean probability for the power pattern to lay within the $k$-th ($k = 1, ..., K$) interval region.

## III. NUMERICAL ASSESSMENT AND PERFORMANCE ANALYSIS

This section is devoted to illustrate the features of the proposed *PIA* method as well as to assess its performance in a representative set of numerical examples concerned with different errors/uncertainties and arrays by proposing some comparisons with Monte Carlo simulations, as well.

The first test case ('*Test Case 1*') refers to a linear array of $N = 16$ elements uniformly-spaced by $d = \frac{\lambda}{2}$ along the $x$-axis. The nominal amplitudes and phases of the array excitations, $\{A_n; n = 1, ..., N\}$ and $\{B_n; n = 1, ..., N\}$, have been set to afford a Taylor pattern steered along broadside with $SLL = -25$ [dB] and $\bar{n} = 3$ (Tab. I). Moreover, the amplifiers and the phase shifters of the array have been assumed to be affected by uniformly-distributed amplitude and phase tolerances with maximum deviations equal to $\xi = 1$ % (i.e., $\xi_n = \xi$; $n = 1, ..., N$) and $\gamma = 3$ [deg] (i.e., $\gamma_n = \gamma$; $n = 1, ..., N$), respectively, with respect to the nominal values in Tab. I. Thus, the actual amplitude and phase excitations turn out to belong to the intervals $\{[A_n]; n = 1, ..., N\}$ and $\{[B_n]; n = 1, ..., N\}$ in Tab. I. Figure 5(a) shows the bounds of the interval power pattern function, $[P(u)]$, computed with the *IA-MS* approach [20], along with the nominal pattern. As it can be observed, although inclusive, the *IA-MS* bounds only provide the information on the worst/largest deviations from the nominal pattern, but nothing more in terms of occurrence of a power pattern value, $\hat{P}(u)$, within such bounds. Of course, the knowledge of bounds that are inclusive is very important for the array designers since they guarantee that no one realization of the power pattern, whatever the tolerance value within the amplitude/phase intervals, can lie outside, but some more could be very helpful in planning the trade-off between costs and robustness to tolerances when building an array for real applications. Towards this purpose, the *PIA* method has been then applied to compute the probabilistic distribution of the values of the actual power pattern within the *IA-MS* bounds, $p_k^P(u)$ ($k = 1, ..., K$) [Figs. 5(*b*)]. By setting the number of strips/region of interest to $K = 5$, it turns out that the one with the highest probability of occurrence is, on average, the $k^* = 3$-th since the mean probability value (13) is equal to $\bar{p}_3 = 26.28$ % (Tab. II), while the less probable region is the first one ($\bar{p}_1 = 9.76$ %). For more immediate and easier inferences, the mean probabilities values of each $k$-th ($k = 1, ..., K$) pattern strip, $\bar{p}_k$, are represented with a grey-scale color within the corresponding bounds in Figs. 5(*c*).

Table I
*Test Case 1* ($N = 16$, $d = \lambda/2$, $\xi = 1$ %, $\gamma = 3$ [DEG]; TAYLOR PATTERN: $SLL = -25$ [DB], $\bar{n} = 3$) - NOMINAL VALUES AND TOLERANCE INTERVALS OF THE AMPLITUDE AND THE PHASE EXCITATIONS OF THE ARRAY.

| $n$ | $A_n$ | $B_n$ [deg] | $[A_n]$ | $[B_n]$ [deg] |
|---|---|---|---|---|
| 1, 16 | 0.365 | 0.0 | [0.362, 0.369] | [−3.0, +3.0] |
| 2, 15 | 0.422 | 0.0 | [0.417, 0.425] | [−3.0, +3.0] |
| 3, 14 | 0.522 | 0.0 | [0.517, 0.527] | [−3.0, +3.0] |
| 4, 13 | 0.646 | 0.0 | [0.640, 0.653] | [−3.0, +3.0] |
| 5, 12 | 0.773 | 0.0 | [0.765, 0.780] | [−3.0, +3.0] |
| 6, 11 | 0.881 | 0.0 | [0.873, 0.890] | [−3.0, +3.0] |
| 7, 10 | 0.960 | 0.0 | [0.950, 0.070] | [−3.0, +3.0] |
| 8, 9 | 1.000 | 0.0 | [0.990, 1.010] | [−3.0, +3.0] |

Table III
*PIA* COMPUTATIONAL COST (*CPU*-TIME).

| | | $K$ | $N$ | $u$ | $N_u$ | $N_v$ | $\gamma$ [deg] | $\Delta t$ [sec] |
|---|---|---|---|---|---|---|---|---|
| *Test Case 1* | Fig. 5(*a*) | − | 16 | [−1, 1] | 501 | 272 | - | 1 |
| | Figs. 5(*b*)-5(*c*) | 5 | 16 | [−1, 1] | 501 | 272 | 3 | 45.14 |
| | Figs. 5(*d*)-5(*e*) | 10 | 16 | [−1, 1] | 501 | 272 | 3 | 47.40 |
| | - | 20 | 16 | [−1, 1] | 501 | 272 | 3 | 50.16 |
| | Figs. 6(*a*)-6(*b*) | 5 | 16 | −0.336 | 1 | 272 | 3 | $9.40 \times 10^{-2}$ |
| | Figs. 6(*c*)-6(*d*) | 10 | 16 | −0.336 | 1 | 272 | 3 | $9.92 \times 10^{-2}$ |
| *Test Case 2* | Figs. 8(*a*)-8(*b*) | 5 | 16 | [−1, 1] | 501 | 112 | 1 | 7.01 |
| | Figs. 5(*b*)-5(*c*) | 5 | 16 | [−1, 1] | 501 | 272 | 3 | 45.14 |
| | Figs. 8(*c*)-8(*d*) | 5 | 16 | [−1, 1] | 501 | 368 | 5 | 102.85 |
| | Figs. 8(*e*)-8(*f*) | 5 | 16 | [−1, 1] | 501 | 720 | 10 | 414.89 |
| *Test Case 3* | Figs. 9(*a*)-9(*b*) | 5 | 8 | [−1, 1] | 251 | 136 | 3 | 13.20 |
| | Figs. 5(*b*)-5(*c*) | 5 | 16 | [−1, 1] | 501 | 272 | 3 | 45.14 |
| | Figs. 9(*c*)-9(*d*) | 5 | 32 | [−1, 1] | 1001 | 544 | 3 | 307.77 |
| | Figs. 9(*e*)-9(*f*) | 5 | 64 | [−1, 1] | 1501 | 1088 | 3 | 1407.46 |

Table II
*Test Case 1* ($N = 16$, $d = \lambda/2$, $\xi = 1$ %, $\gamma = 3$ [DEG]; TAYLOR PATTERN: $SLL = −25$ [DB], $\bar{n} = 3$) - MEAN PROBABILITY VALUES $\{\bar{p}_k; k = 1, ..., K\}$ WHEN $K = 5$ AND $K = 10$.

| $K = 5$ | | $K = 10$ | | | |
|---|---|---|---|---|---|
| $k$ | $\bar{p}_k$ [%] | $k$ | $\bar{p}_k$ [%] | $k$ | $\bar{p}_k$ [%] |
| 1 | 9.76 | 1 | 2.84 | 2 | 6.92 |
| 2 | 21.59 | 3 | 9.84 | 4 | 11.25 |
| 3 | 26.28 | 5 | 12.81 | 6 | 13.47 |
| 4 | 26.19 | 7 | 13.51 | 8 | 12.68 |
| 5 | 16.18 | 9 | 10.41 | 10 | 5.77 |

Depending on the user-required spatial resolution and thanks to the computational efficiency of the *PIA* (Tab. III), there are no problems in having a more detailed power pattern tolerance distribution by considering narrower/finer regions (i.e., greater values of $K$), the *CPU*-time being almost independent on $K$ (Tab. III) thanks to the closed-form expressions detailed in *Appendix I* (see *Case 1* - *Case 4*). As an example, the step-wise distribution of the power pattern occurrence when $K = 10$ is reported in Fig. 5(*d*), while the corresponding mean probability values are given Tab. II and displayed in Fig. 5(*e*). As it can be observed, the comparisons between the probability plots in Fig. 5(*d*) and Fig. 5(*b*) as well as those in Fig. 5(*e*) and Fig. 5(*c*) point out the higher resolution in the definition of the power pattern *PDF* within the *IA-MS* bounds, $P_{IA-MS}^{inf}(u)$ and $P_{IA-MS}^{sup}(u)$, thanks to the increased number of $K$ partitions. On the other hand, the values of $p_k^P(u)$ ($k = 1, ..., K$) reduce as $K$ increases both on average [see the dynamic of the grey-scale in Fig. 5(*e*) vs. Fig. 5(*c*) - Tab. II] and whatever the angular direction (i.e., $u$ value) being by definition $\sum_{k=1}^{K} p_k^P(u) = 1$ as well as $\sum_{k=1}^{K} \bar{p}_k = 1$. Indeed, since $w_k \rfloor_{K=5} = 2 \times w_k \rfloor_{K=10}$, it turns out that $p_h^P(u) \rfloor_{K=5} = p_{2h-1}^P(u) \rfloor_{K=10} + p_{2h}^P(u) \rfloor_{K=10}$ ($h = 1, ..., K$; $K = 5$) [Fig. 5(*b*) vs. Fig. 5(*d*)] and $\bar{p}_h \rfloor_{K=5} = \bar{p}_{2h-1} \rfloor_{K=10} + \bar{p}_{2h} \rfloor_{K=10}$ ($h = 1, ..., K$; $K = 5$) (Tab. II).

Next, the results from *PIA* have been compared with those from the Monte Carlo method when evaluating $\aleph = 10^6$ realizations/configurations of the excitation tolerances within the ranges $\{[W_n]; n = 1, ..., N\}$ defined by the deviations $\xi = 1$ % and $\gamma = 3$ [deg]. For illustrative purposes, Fig. 6(*a*) and Fig. 6(*c*) give the representation in the complex plane of the *AF* interval function sampled at the angular direction $u_0 = −0.336$ (i.e., the surface mapping the complex interval $[AF(u_0)]$ in the complex plane) and predicted by the *IA-MS* method. Moreover, the probability that the actual power pattern value lies within the $k$-th region ($k = 1, ..., K$), $p_k^P(u_0)$[3], computed by the *PIA* when $K = 5$ [Fig. 6(*a*)] and $K = 10$ [Fig. 6(*c*)] is reported in color within the $[AF(u_0)]$ area, as well. As expected, $p_h^P \rfloor_{K=5} = p_{2h-1}^P \rfloor_{K=10} + p_{2h}^P \rfloor_{K=10}$ (Tab. IV) being $[P_h] \rfloor_{K=5} = [P_{2h-1}] \rfloor_{K=10} + [P_{2h}] \rfloor_{K=10}$ ($h = 1, ..., K$; $K = 5$). Concerning the comparative study, the plot of the *PDF* of the Monte Carlo simulations is reported in both Fig. 6(*b*) and Fig. 6(*d*) along with the color-map representation of the $K$ probability regions, $[P_k]$, and their probability values, $p_k^P$ ($k = 1, ..., K$). More in detail, each $k$-th ($k = 1, ..., K$) vertical strip extends from the infimum, $P_k^{inf}$ ($P_k^{inf} \triangleq \left| AF_k^{inf}(u_0) \right|^2$), to the supremum, $P_k^{sup}$ ($P_k^{sup} \triangleq |AF_k^{sup}(u_0)|^2$), of the corresponding *PIA*-derived interval power pattern region. It is worth noticing that the bell of the Gaussian-like Monte Carlo *PDF* is localized within the most probable $k$-th [$k^* = 3$ - Fig. 6(*b*); $k^* = 6$ - Fig. 6(*d*)] region predicted with the *PIA* method [Figs. 6(*b*)-6(*d*)] to confirm the reliability of the proposed probabilistic method. On the other hand, the interested reader needs to take into account that the *PIA* prediction has been yielded with just a single analytic computation, instead of running $\aleph$ times the computation of the power pattern, carried out in only few milliseconds ($\Delta t = 9.40 \times 10^{-2}$ [sec] - $K = 5$; $\Delta t = 9.92 \times 10^{-2}$ [sec] - $K = 10$; Tab. III) to derive an inclusive result, while the non-exhaustive Monte Carlo estimation run for some hours.

---

[3] The following simplification of the notation is used hereinafter: $p_k^P \leftarrow p_k^P(u_0)$ and $[P_k] \leftarrow [P_k(u_0)]$ being $p_k^P(u_0) \triangleq \Pr\left\{\hat{P}(u_0) \in [P_k(u_0)]\right\}$

Table IV
*Test Case 1* ($N = 16$, $d = \lambda/2$, $\xi = 1$ %, $\gamma = 3$ [DEG]; TAYLOR PATTERN: $SLL = -25$ [DB], $\bar{n} = 3$; $u_0 = -0.336$) - POWER PATTERN PROBABILITY REGIONS $\{[P_k]; k = 1, ..., K\}$ AND THEIR PROBABILITY VALUES $\{p_k^P; k = 1, ..., K\}$ WHEN $K = 5$ AND $K = 10$.

| $K = 5$ | | | $K = 10$ | | | | | |
|---|---|---|---|---|---|---|---|---|
| $k$ | $p_k^P$ [%] | $[P_k]$ [dB] | $k$ | $p_k^P$ [%] | $[P_k]$ [dB] | $k$ | $p_k^P$ [%] | $[P_k]$ [dB] |
| 1 | 7.46 | $[-54.98, -34.76]$ | 1 | 2.15 | $[-54.98, -39.97]$ | 2 | 5.31 | $[-39.97, -34.76]$ |
| 2 | 19.59 | $[-34.76, -29.17]$ | 3 | 8.31 | $[-34.76, -31.53]$ | 4 | 11.28 | $[-31.53, -29.17]$ |
| 3 | 28.30 | $[-29.17, -25.80]$ | 5 | 13.87 | $[-29.17, -27.33]$ | 6 | 14.43 | $[-27.33, -25.80]$ |
| 4 | 27.41 | $[-25.80, -23.38]$ | 7 | 13.91 | $[-25.80, -24.51]$ | 8 | 13.50 | $[-24.51, -23.38]$ |
| 5 | 17.25 | $[-23.38, -21.49]$ | 9 | 12.49 | $[-23.38, -22.38]$ | 10 | 4.76 | $[-22.38, -21.49]$ |

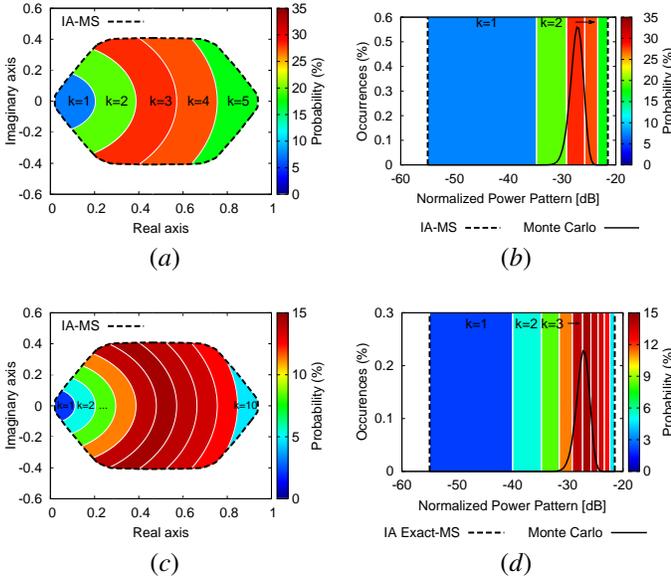

Figure 6. *Test Case 1* ($N = 16$, $d = \lambda/2$, $\xi = 1$ %, $\gamma = 3$ [deg]; Taylor pattern: $SLL = -25$ [dB], $\bar{n} = 3$; $u_0 = -0.336$) - Plot of (*a*)(*c*) the complex-valued *IA-MS* array factor interval, $[AF(u_0)]$, partitioned into $K$ power pattern probability regions $\{[P_k(u_0)]; k = 1, ..., K\}$ with the color-level representation of the probability values $\{p_k^P(u_0); k = 1, ..., K\}$ and (*b*)(*d*) comparison between the *PDF*s from the Monte Carlo simulations and derived with the *PIA* when (*a*)(*b*) $K = 5$ and (*c*)(*d*) $K = 10$.

Table V
*Test Case 1* ($N = 16$, $d = \lambda/2$, $\xi = 1$ %, $\gamma = 3$ [DEG]; TAYLOR PATTERN: $SLL = -25$ [DB], $\bar{n} = 3$; $K = 5$) - SIDELOBE LEVEL INTERVALS $\{[SLL_k]; k = 1, ..., K\}$ AND POWER PATTERN PEAK INTERVALS $\{[\Gamma_k]; k = 1, ..., K\}$ ALONG WITH ITS PROBABILITIES $\{p_k^\Gamma; k = 1, ..., K\}$.

| $k$ | $[SLL_k]$ [dB] | $p_k^\Gamma$ [%] | $[\Gamma_k]$ [dB] |
|---|---|---|---|
| 1 | $[-37.08, -30.25]$ | 18.11 | $[-0.099, -0.062]$ |
| 2 | $[-30.43, -26.53]$ | 20.35 | $[-0.062, -0.024]$ |
| 3 | $[-26.71, -23.93]$ | 20.44 | $[-0.024, 0.013]$ |
| 4 | $[-24.12, -21.93]$ | 20.52 | $[0.013, 0.050]$ |
| 5 | $[-22.12, -20.31]$ | 20.58 | $[0.050, 0.087]$ |
| $IA - MS$ | $[-37.08, -20.31]$ | — | $[-0.099, 0.087]$ |

A very fruitful and free by-product of the *PIA* analytic prediction of the power pattern tolerance distribution is the easy derivation of the *PDF* of the pattern features. With reference to the intervals of the $SLL$ and of the beam pattern peak $\Gamma$, whose endpoints are [18]

$$SLL^{inf/sup} \triangleq P^{inf/sup}(u_{max}) - P^{sup/inf}(u_{SLL}) \quad (14)$$

where $u_{max}$ is the steering direction and $u_{SLL} \triangleq \arg\{\max_{u \notin \Omega}\{P(u)\}\}$, $\Omega$ being the sidelobe-region outside the mainbeam, and

$$\Gamma^{inf/sup} \triangleq P^{inf/sup}(u_{max}) \quad (15)$$

respectively, the same definitions can be now extended to each $k$-th ($k = 1, ..., K$) probability region of the power pattern. Accordingly, let us define

$$SLL_k^{inf/sup} \triangleq P^{inf/sup}(u_{max}) - P_k^{sup/inf}(u_{SLL(k)}) \quad (16)$$

where $u_{SLL(k)} \triangleq \arg\{\max_{u \notin \Omega_k}\{\hat{P}(u)\}\}$, while

$$\Gamma_k^{inf/sup} \triangleq P_k^{inf/sup}(u_{max}). \quad (17)$$

As for the corresponding *PDF*s, the probability of $\Gamma$ ($p_k^\Gamma \triangleq \Pr\{\hat{\Gamma} \in [\Gamma_k]\}$) is given by $p_k^\Gamma = \Pr\{\hat{P}(u_{max}) \in [P_k(u_{max})]\}$, while the occurrence distribution of the *SLL*, $p_k^{SLL} \triangleq \Pr\{\widehat{SLL} \in [SLL_k]\}$, is approximated with the mean probability of the $k$-th ($k = 1, ..., K$) power pattern strip (i.e., $p_k^{SLL} \approx \bar{p}_k$) since $p_k^{SLL} \to \bar{p}_k$ ($K \to \infty$).

By applying (16) and (17) to the *IA-MS* intervals $[SLL]^{IA-MS} = [-37.08, -20.31]$ [dB] and $[\Gamma]^{IA-MS} = [-0.099, 0.087]$ [dB] (Tab. V) of the *Test Case 1* (i.e., $N = 16$, $d = \lambda/2$, $\xi = 1$ %, $\gamma = 3$ [deg]; Taylor pattern: $SLL = -25$ [dB], $\bar{n} = 3$), the $K$ intervals of the pattern features and their probabilities turn out to be those in Tab. V ($K = 5$). As it can be noticed, both the peak power and the *SLL* intervals perfectly fill the *IA-MS* bounds (i.e., $SLL_K^{sup} = SLL_{IA-MS}^{sup}$ and $SLL_1^{inf} = SLL_{IA-MS}^{inf}$; $\Gamma_K^{sup} = \Gamma_{IA-MS}^{sup}$ and $\Gamma_1^{inf} = \Gamma_{IA-MS}^{inf}$ - Tab. V), but only those of $\Gamma$ are adjacent [i.e., $\Gamma_h^{sup} = \Gamma_{h+1}^{inf}$ ($h = 1, ..., K-1$)] and mutually exclusive (i.e., $[\Gamma_h] \cap [\Gamma_k] = \emptyset$; $h, k = 1, ..., K$; $h \neq k$) since $u_{max}$ is a fixed direction in (17), while the angular direction of the highest sidelobe in the $k$-th region, $u_{SLL(k)}$, in (16) can vary. On the other hand, a key outcome is that Fig. 7 and the values in Tab. V give faithful indications on the most/less probable values of the actual *SLL* [Fig. 7(a)] and $\Gamma$ [Fig. 7(b)] within the *IA-MS* bounds. For instance, the *SLL* most probably has a value in the range $SLL_k^{inf}\big|_{k=3} \leq SLL \leq SLL_k^{sup}\big|_{k=3}$ ($SLL_k^{inf}\big|_{k=3} = -26.71$ [dB] and $SLL_k^{sup}\big|_{k=3} = -23.93$ [dB]) since $p_3^{SLL} \geq p_h^{SLL}$ ($h = 1, ..., K$) being $p_3^{SLL} = 26.28$ %, while the lowest probability arises for the ($k = 1$)-th interval of the admissible *SLL* values [Fig. 7(a)] being $p_1^{SLL} = 9.76$ %. Such an analysis clearly highlights that, thanks to the *PIA* technique, more information than that coded into the *IA-MS* endpoints/bounds can be drawn from the knowledge of $[AF(u)]$. As a matter of fact and unlike *PIA*, standard *IA*-based methods only provides the worst case bounds, which are generally characterized by a low or very-low probability of occurrence.

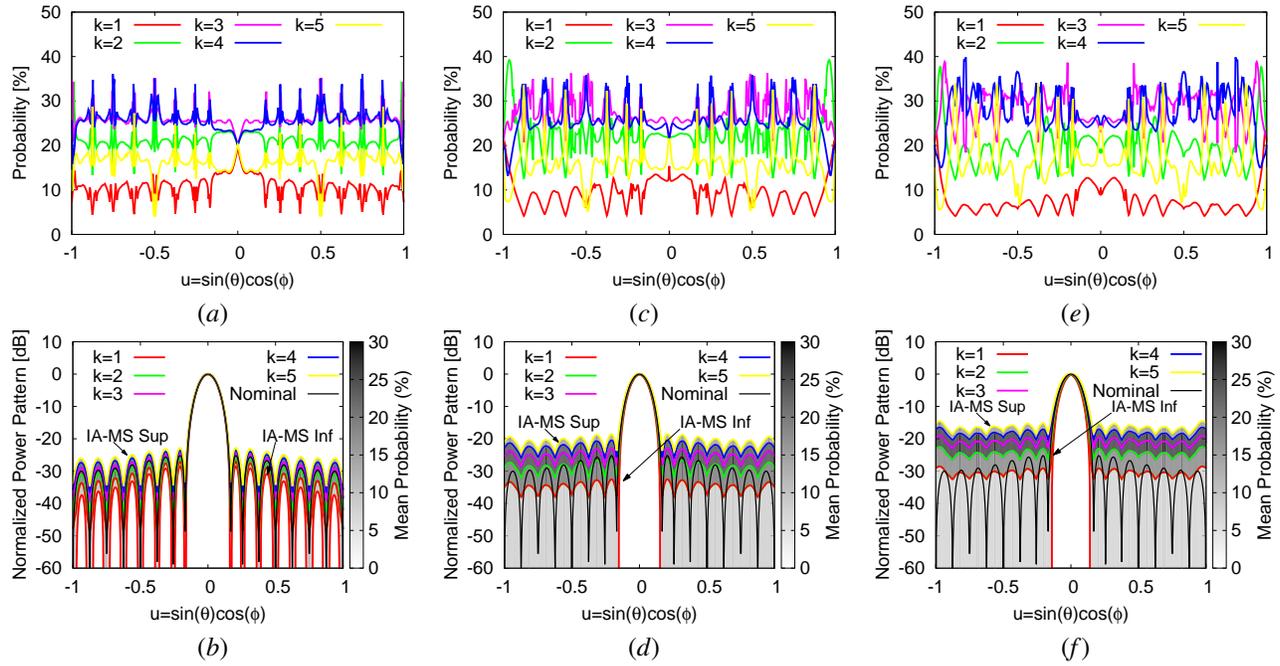

Figure 8. *Test Case 2* ($N = 16$, $d = \lambda/2$, $\xi = 1$ %; Taylor pattern: $SLL = -25$ [dB], $\bar{n} = 3$; $K = 5$) - Plot of (a)(c)(e) the probability functions $\{p_k^P(u); k = 1, ..., K\}$, and (b)(d)(f) the $K$ power pattern probability regions $\{[P_k(u)]; k = 1, ..., K\}$ with the grey-level representation of the mean probability values $\{\bar{p}_k; k = 1, ..., K\}$ when (a)(b) $\gamma = 1$ [deg], (c)(d) $\gamma = 5$ [deg], and (e)(f) $\gamma = 10$ [deg].

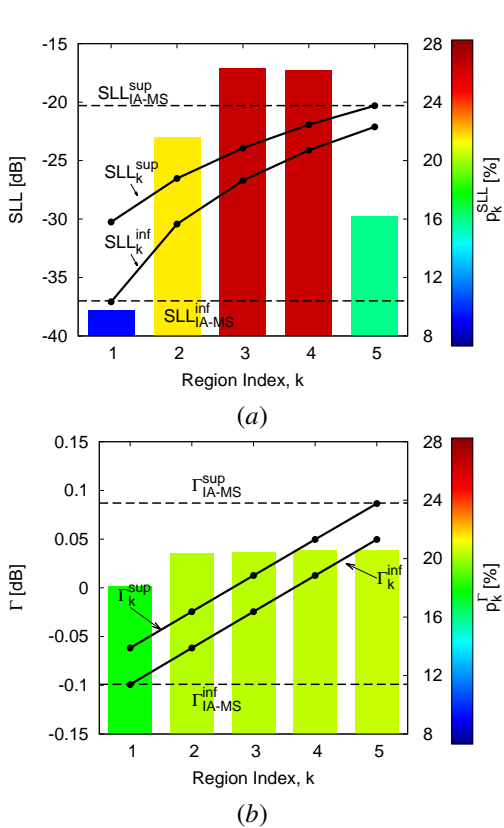

Figure 7. *Test Case 1* ($N = 16$, $d = \lambda/2$, $\xi = 1$ %, $\gamma = 3$ [deg]; Taylor pattern: $SLL = -25$ [dB], $\bar{n} = 3$; $K = 5$) - Bounds of (a) the sidelobe level intervals $\{[SLL_k]; k = 1, ..., K\}$ and (b) the power pattern peak intervals $\{[\Gamma_k]; k = 1, ..., K\}$ along with their probability values, $\{p_k^{SLL}; k = 1, ..., K\}$ and $\{p_k^{\Gamma}; k = 1, ..., K\}$.

In order to further assess the *PIA* method as well as its performance, different values of the uniform error tolerances on the phase excitations have been considered ('*Test Case 2*'), while keeping the uniform amplitude deviation fixed to $\xi = 1$ % as well as the array and the radiated pattern (Taylor beam: $SLL = -25$ [dB] and $\bar{n} = 3$ - Tab. I). The probabilistic distributions of the values of the actual power pattern within the *IA-MS* bounds, $p_k^P(u)$, $u \in [-1, 1]$ ($k = 1, ..., K$; $K = 5$), are shown on the left column of Fig. 8 when $\gamma = 1$ [deg] [Fig. 8(a)], $\gamma = 5$ [deg] [Fig. 8(c)], and $\gamma = 10$ [deg] [Fig. 8(e)]. Moreover, the mean probability value (13) of each $k$-th ($k = 1, ..., K$) pattern strip, $\bar{p}_k \rfloor_\gamma$ ($\gamma \in \{1, 5, 10\}$ [deg]), is reported with the grey-level representation in Fig. 8(b) ($\gamma = 1$ [deg]), Fig. 8(d) ($\gamma = 5$ [deg]), and Fig. 8(f) ($\gamma = 10$ [deg]) as well as in Tab. VI where, for completeness, the intervals and the probabilities of the $SLL$, $\{[SLL_k]; k = 1, ..., K\}$, and of the power pattern peak, $\{[\Gamma_k]; k = 1, ..., K\}$ are given, as well. Both figures and tables indicate that, on average and analogously to the first test case, the most probable region turns out to be the central one ($k^* = 3$), whatever the phase tolerance $\gamma$ ($\bar{p}_3 \rfloor_{\gamma=1 \text{ [deg]}} = 26.00$ %, $\bar{p}_3 \rfloor_{\gamma=5 \text{ [deg]}} = 27.19$ %, and $\bar{p}_3 \rfloor_{\gamma=10 \text{ [deg]}} = 28.01$ % - Tab. VI). Moreover, the regions close to the *IA-MS* bounds (i.e., $k = 1$ and $k = K$) are, once again, the less probable ones. Such an outcome further points out a key feature of the *PIA* as well as its advantage over standard, even though inclusive, *IA*-based tools for tolerance analysis, which are unable to give information on the occurrences of the actual power patter within the interval bounds.

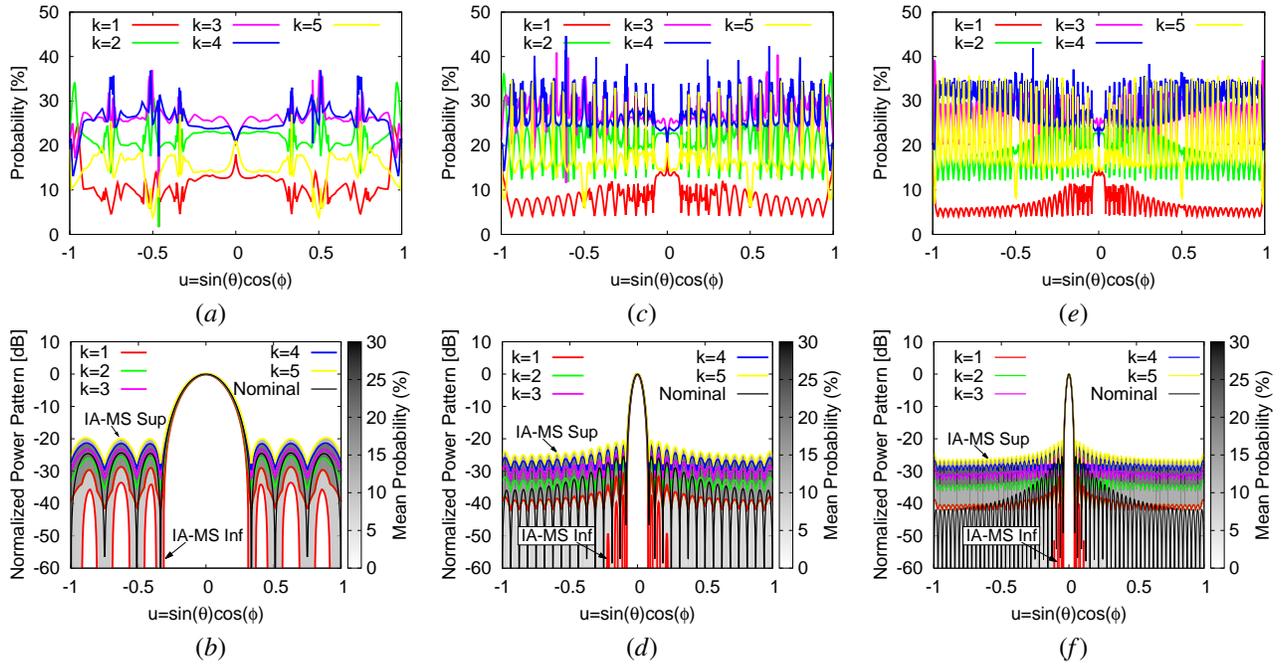

Figure 9. Test Case 3 ($d = \lambda/2$, $\xi = 1$ %, $\gamma = 3$ [deg]; Taylor pattern: $SLL = -25$ [dB], $\bar{n} = 3$; $K = 5$) - Plot of (a)(c)(e) the probability functions $\{p_k^P(u); k = 1, ..., K\}$, and (b)(d)(f) the $K$ power pattern probability regions $\{[P_k(u)]; k = 1, ..., K\}$ with the grey-level representation of the mean probability values $\{\bar{p}_k; k = 1, ..., K\}$ for an array of (a)(b) $N = 8$, (c)(d) $N = 32$, and (e)(f) $N = 64$ elements.

As for the computational costs, the values in Tab. III indicate that the *CPU*-time $\Delta t$ depends on the width of the phase tolerance, $\gamma$, since the extension of $\mathcal{A}\{[AF(u)]\}$ grows proportionally. Indeed, more vertexes are needed for the *IA-MS* operations [20] in the case of larger deviations from the nominal excitations, then more triangles are used to partition the interval $[AF(u)]$ for computing $p_k(u)$ ($k = 1, ..., K$) [Fig. A.2]. For the sake of completeness, the values of the number of vertexes ($N_v$) used for each $u$ sample and of the number of samples ($N_u$) used to discretize the interval $u = [-1, 1]$ are reported in Tab. III.

The last example ('*Test Case 3*') is aimed at analyzing the *PIA* method when dealing with different linear arrays. Towards this end, different array sizes, ranging from $N = 8$ up to $N = 64$ elements, have been considered by still keeping an uniform half-wavelength inter-element spacing ($d = \frac{\lambda}{2}$) and the uniform excitation tolerances to $\xi = 1$ % and $\gamma = 3$ [deg]. When applying the *PIA* method ($K = 5$) to the *IA-MS* bounds of $[AF(u)]$, the *PDF* of the power pattern (Fig. 9) has been computed with a non-optimized code on a standard laptop PC having $2.40 GHz$ CPU and $6 GB$ RAM in an amount of time $\Delta t$ that significantly depends on $N$ (Tab. III) due to the *IA-MS* computation of (6). Indeed, since the *IA-MS* calculation of $[AF(u)]$ requires the sum of $N$ interval phasors for each $u$-direction and the sampling rate of the angular range $u \in [-1, 1]$ is proportional to $N$, the number of vertexes of the surface mapping the $[AF(u)]$ function in the complex plane grows with $\mathcal{O}(N^2)$. This implies a higher cost of the *PIA* prediction when increasing the array size (i.e., $N$) with respect to the widening of the $\gamma$ interval (Tab. III). As for the plots in Fig. 9, it can be observed that the variations of $p_k^P(u)$ ($k = 1, ..., K$) versus the angular direction $u$, within the range $u \in [-1, 1]$, increase as the array enlarges. For example, let us compare the behavior of $p_1^P(u)$ in Fig. 9($a$) and Fig. 9($e$). Such an outcome is expected since the condition $p_k^P(u) \approx p_k^P(u + \Delta u)$ ($\Delta u$ being a very small increment, $\Delta u \ll 1$) holds true when $\hat{P}(u) \approx \hat{P}(u + \Delta u)$, but this latter relation turns out to be more and more unfulfilled as $N$ grows. Concerning the most probable power pattern region (right column in Fig. 9 and Tab. VII), the highest value of $\bar{p}_k$ occurs in the center of the partitioning of the *IA-MS* bounds (i.e., $k^* = 3$) when $N \le 32$, while the actual power pattern value tends to fall nearer to the *IA-MS* upper bound ($k^* \to K$) for larger arrays (e.g., $k^*|_{N=64} = 4$).

Table VI
Test Case 2 ($N = 16$, $d = \lambda/2$, $\xi = 1$ %; Taylor pattern: $SLL = -25$ [dB], $\bar{n} = 3$; $K = 5$) - Sidelobe level intervals $\{[SLL_k]; k = 1, ..., K\}$ and power pattern peak intervals $\{[\Gamma_k]; k = 1, ..., K\}$ along with their probabilities $\{p_k^{SLL}; k = 1, ..., K\}$ and $\{p_k^\Gamma; k = 1, ..., K\}$ when $\gamma \in \{1, 5, 10\}$ [deg].

| $k$ | $\bar{p}_k$ [%] | $[SLL_k]$ [dB] | $p_k^\Gamma$ [%] | $[\Gamma_k]$ [dB] |
|---|---|---|---|---|
| | | $\gamma = 1$ [deg] | | |
| 1 | 10.78 | $[-28.68, -27.00]$ | 19.64 | $[-0.087, -0.053]$ |
| 2 | 21.27 | $[-27.18, -25.72]$ | 19.98 | $[-0.053, -0.018]$ |
| 3 | 26.00 | $[-25.89, -24.60]$ | 20.06 | $[-0.018, 0.017]$ |
| 4 | 25.77 | $[-24.77, -23.61]$ | 20.14 | $[0.017, 0.052]$ |
| 5 | 16.19 | $[-23.78, -22.72]$ | 20.18 | $[0.052, 0.087]$ |
| $IA - MS$ | – | $[-28.68, -22.72]$ | – | $[-0.087, 0.087]$ |
| | | $\gamma = 5$ [deg] | | |
| 1 | 9.38 | $[-\infty, -32.41]$ | 15.36 | $[-0.120, -0.079]$ |
| 2 | 21.90 | $[-32.61, -26.38]$ | 21.02 | $[-0.079, -0.037]$ |
| 3 | 27.19 | $[-26.59, -22.86]$ | 21.12 | $[-0.037, 0.004]$ |
| 4 | 25.83 | $[-23.07, -20.36]$ | 21.22 | $[0.004, 0.046]$ |
| 5 | 15.70 | $[-20.57, -18.42]$ | 21.28 | $[0.046, 0.087]$ |
| $IA - MS$ | – | $[-\infty, -18.42]$ | – | $[-0.120, 0.087]$ |
| | | $\gamma = 10$ [deg] | | |
| 1 | 7.64 | $[-\infty, -27.66]$ | 10.51 | $[-0.220, -0.158]$ |
| 2 | 19.83 | $[-27.97, -21.64]$ | 19.22 | $[-0.158, -0.096]$ |
| 3 | 28.01 | $[-21.95, -18.12]$ | 23.19 | $[-0.096, -0.034]$ |
| 4 | 27.97 | $[-18.43, -15.62]$ | 23.47 | $[-0.034, 0.026]$ |
| 5 | 16.55 | $[-15.93, -13.68]$ | 23.61 | $[0.026, 0.087]$ |
| $IA - MS$ | – | $[-\infty, -13.68]$ | – | $[-0.220, 0.087]$ |

Table VII
Test Case 3 ($d = \lambda/2$, $\xi_n = 1\%$, $\gamma = 3$ [DEG]; TAYLOR PATTERN: $SLL = -25$ [DB], $\bar{n} = 3$; $K = 5$) - SIDELOBE LEVEL INTERVALS $\{[SLL_k]; k = 1, ..., K\}$ AND POWER PATTERN PEAK INTERVALS $\{[\Gamma_k]; k = 1, ..., K\}$ ALONG WITH THEIR PROBABILITIES $\{p_k^{SLL}; k = 1, ..., K\}$ AND $\{p_k^\Gamma; k = 1, ..., K\}$ WHEN $N \in \{8, 32, 64\}$.

| $k$ | $\bar{p}_k$ [%] | $[SLL_k]$ [dB] | $p_k^\Gamma$ [%] | $[\Gamma_k]$ [%] |
|---|---|---|---|---|
| | | $N = 8$ | | |
| 1 | 11.28 | $[-33.61, -28.58]$ | 18.01 | $[-0.099, -0.062]$ |
| 2 | 21.92 | $[-28.76, -25.47]$ | 20.37 | $[-0.062, -0.024]$ |
| 3 | 26.13 | $[-25.65, -23.18]$ | 20.46 | $[-0.024, 0.013]$ |
| 4 | 25.62 | $[-23.37, -21.24]$ | 20.55 | $[0.013, 0.050]$ |
| 5 | 15.05 | $[-21.43, -19.55]$ | 20.61 | $[0.050, 0.087]$ |
| $IA - MS$ | − | $[-33.61, -19.55]$ | − | $[-0.099, 0.087]$ |
| | | $N = 32$ | | |
| 1 | 8.17 | $[-37.63, -30.54]$ | 18.06 | $[-0.099, -0.062]$ |
| 2 | 20.83 | $[-30.73, -26.74]$ | 20.36 | $[-0.062, -0.024]$ |
| 3 | 27.07 | $[-26.93, -24.11]$ | 20.45 | $[-0.024, 0.013]$ |
| 4 | 27.05 | $[-24.29, -22.09]$ | 20.54 | $[0.013, 0.050]$ |
| 5 | 16.88 | $[-22.28, -20.45]$ | 20.59 | $[0.050, 0.087]$ |
| $IA - MS$ | − | $[-37.63, -20.45]$ | − | $[-0.099, 0.087]$ |
| | | $N = 64$ | | |
| 1 | 6.56 | $[-37.80, -30.62]$ | 18.12 | $[-0.089, -0.054]$ |
| 2 | 18.01 | $[-30.80, -26.79]$ | 20.35 | $[-0.054, -0.019]$ |
| 3 | 26.87 | $[-26.98, -24.15]$ | 20.43 | $[-0.019, 0.017]$ |
| 4 | 29.48 | $[-24.33, -22.12]$ | 20.52 | $[0.017, 0.052]$ |
| 5 | 19.08 | $[-22.31, -20.47]$ | 20.58 | $[0.052, 0.087]$ |
| $IA - MS$ | − | $[-37.80, -20.47]$ | − | $[-0.089, 0.087]$ |

For the sake of completeness, the $K$ intervals of the pattern features (i.e., *SLL* and $\Gamma$) along with their probabilities are reported in Tab. VII, as well.

## IV. CONCLUSION

A probabilistic *IA*-based approach has been proposed to deal with the tolerance analysis of linear phased arrays affected by arbitrary, but bounded, uncertainties/errors in both amplitude and phase array excitations. More specifically, an analytic method has been described to derive the closed-form expression of the probability distribution of the power pattern tolerance within the reliable, yet inclusive, interval bounds yielded with the *IA-MS* technique [20].

From a methodological viewpoint, the main novelties and advantages, over the existing state-of-the-art techniques, of the proposed *PIA* method can be summarized in the following ones:

- the inclusion, for the first time to the best of the authors' knowledge, of a probabilistic information within the worst-case *IA* bounds to provide the antenna engineers a more informative, but still inclusive and analytic, prediction on the expected *PA* performance;
- the development of a computationally-efficient and exact (i.e., without any approximation) technique for extracting the probabilistic information coded into the representation in the complex plane of the envelope of the complex interval function of the array radiation pattern.

The numerical assessment has provided evidence of the effectiveness, the reliability, and the computational efficiency of the *PIA* method in dealing with different error tolerances in the array excitations as well as array dimensions.

Future works, beyond the scope of this paper, will be aimed at extending the proposed method to large arrays, also considering planar and conformal layouts with uniform and non-uniform/sparse element arrangements. Moreover, further advances will also consider the application of the *PIA* method to the analysis of other pattern features (e.g., directivity, gain) and its integration into an optimization tool for the robust *PA* synthesis.

*Appendix I*

The term $\mathcal{A}\{[AF_k(u)]\}$ at the numerator of (11) is the portion of $[AF(u)]$ inside the $k$-th ring $[G_k(u)]$, $\mathcal{A}\{[AF(u)] \cap C_k(u)\}$ [Fig. A.1($c$)]

$$\mathcal{A}\{[AF_k(u)]\} \triangleq \mathcal{A}\{[AF(u)] \cap [G_k(u)]\} \quad (18)$$

given by the difference between the sub-region of $[AF(u)]$ inside the upper circle $C_{k+1}(u)$, $\mathcal{A}\{[AF(u)] \cap C_{k+1}(u)\}$ [Fig. A.1($a$)] and that bounded by the lower circle $C_k(u)$, $\mathcal{A}\{[AF(u)] \cap C_k(u)\}$ [Fig. A.1($b$)]

$$\mathcal{A}\{[AF_k(u)]\} = \mathcal{A}\{[AF(u)] \cap C_{k+1}(u)\} \\ - \mathcal{A}\{[AF(u)] \cap C_k(u)\}. \quad (19)$$

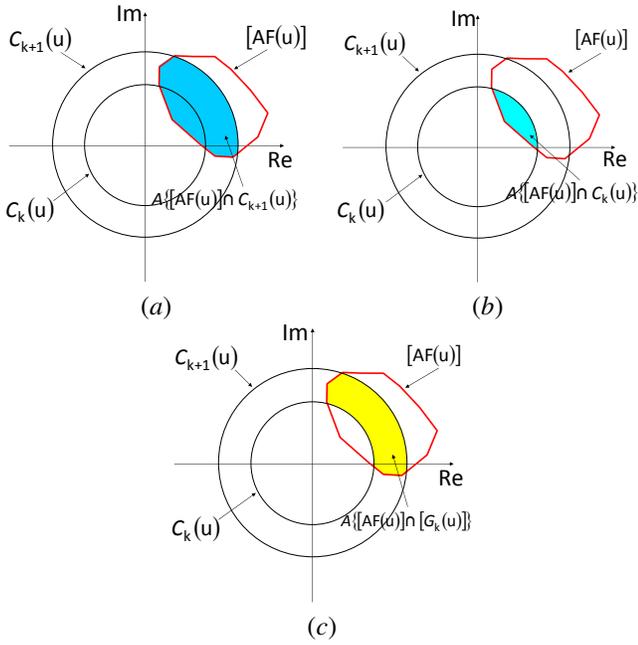

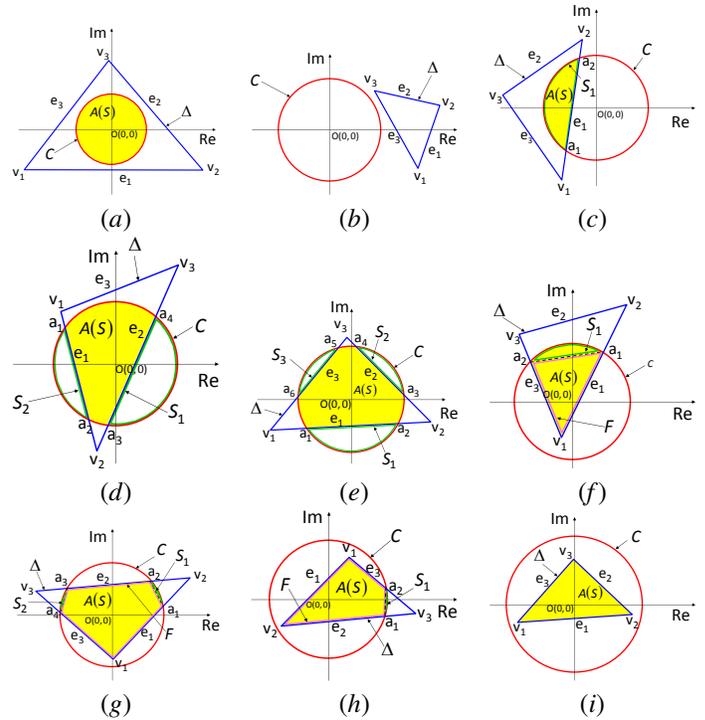

Figure A.1. Illustrative sketches of (*a*) the sub-region of $[AF(u)]$ inside the upper circle $C_{k+1}(u)$ with area $\mathcal{A}\{[AF(u)] \cap C_{k+1}(u)\}$, (*b*) the sub-region of $[AF(u)]$ inside the upper circle $C_k(u)$ with area $\mathcal{A}\{[AF(u)] \cap C_k(u)\}$, and (*c*) the portion of $[AF(u)]$ inside the $k$-th ring $[G_k(u)]$ having area

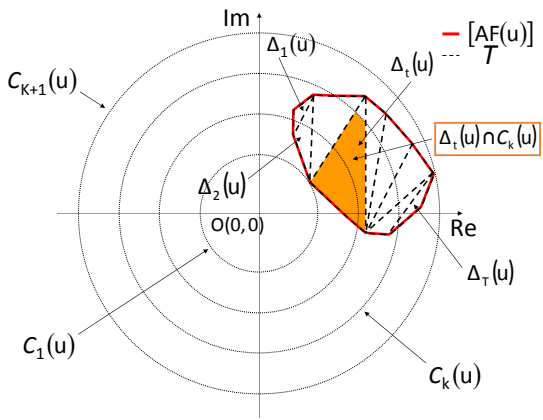

Figure A.2. Illustrative sketch in the complex plane of the partition of $[AF(u)]$ into $T$ triangles, $\{\Delta_t(u); t = 1, ..., T\}$.

In order to compute the intersection areas between $[AF(u)]$ and the circles $C_{k+1}(u)$ and $C_k(u)$, a triangulation of $[AF(u)]$ is exploited by applying the Delaunay algorithm [40][41] to the vertexes used by the *IA-MS* operations [20] to determine (6). More specifically, the interval $[AF(u)]$ is partitioned into $T$ triangles, $\{\Delta_t(u); t = 1, ..., T\}$ (Fig. A.2), and the generic $h$-th ($h = 1, ..., K+1$) term $\mathcal{A}\{[AF(u)] \cap C_h(u)\}$ is then yielded as the sum of all the common regions between the $T$ triangles and the circle $C_h(u)$

$$\mathcal{A}\{[AF(u)] \cap C_h(u)\} = \sum_{t=1}^{T} \mathcal{A}\{C_h(u) \cap \Delta_t(u)\} \quad (20)$$

Figure A.3. Illustrative sketches for the computation of the area of the crossing region $\mathcal{S}$ ($\mathcal{S} \triangleq C \cap \Delta$) between a generic circle $C$ ($C \in \{C_h(u); h = 1, ..., K+1\}$) of radius $r$ ($r \in \{r_h; h = 1, ..., K+1\}$) and a triangle $\Delta$ ($\Delta \in \{\Delta_t(u); t = 1, ..., T\}$) of anti-clockwise ordered vertexes $\{v_j ; j = 1, ..., 3\}$ with edges $\{e_j; j = 1, ..., 3\}$. *Case 1* - All vertexes of $\Delta$ are outside the circle ($v_j \notin C$; $j = 1, ..., 3$) and there are not edge intersections with $C$ ($e_j \cap C = \emptyset$; $j = 1, ..., 3$) (*a*)(*b*). *Case 2* - All vertexes of $\Delta$ are outside the circle $C$ and there is/are (*c*) one or (*d*) two or (*e*) three edges intersections. *Case 3* - (*f*)(*h*) One or (*g*) two vertexes of $\Delta$ fall within the circle $C$. *Case 4* - All three vertexes of $\Delta$ lie within the circle $C$ (*i*).

where $\mathcal{A}\{C_h(u) \cap \Delta_t(u)\}$ is calculated by considering the mutual position of the $t$-th triangle $\Delta_t(u)$ and the circumference $C_h(u)$ and according to the rules pictorially summarized in Fig. A.3. As for these latter, the guideline for computing the crossing area $\mathcal{S}$ ($\mathcal{S} \triangleq C \cap \Delta$) between a generic circle $C$ ($C \in \{C_h(u); h = 1, ..., K+1\}$) of radius $r$ ($r \in \{r_h; h = 1, ..., K+1\}$) and a triangle $\Delta$ ($\Delta \in \{\Delta_t(u); t = 1, ..., T\}$) of anti-clockwise ordered vertexes $\{v_j ; j = 1, ..., 3\}$ with edges $\{e_j; j = 1, ..., 3\}$, is that of checking the relative positions of the vertexes and edges with respect to $C$. More specifically, depending on the number of vertexes that lie within the circle $C$ and the number of intersections between $C$ and $\Delta$, the following cases arise:

- **Case 1** - If all the vertexes are outside the circle ($v_j \notin C$; $j = 1, ..., 3$) and there are not edge intersections with $C$ ($e_j \cap C = \emptyset$; $j = 1, ..., 3$) [Figs. A.3(*a*)-A.3(*b*)], the value of $\mathcal{A}(\mathcal{S})$ is equal to: (*a*) the whole circle surface $\mathcal{A}(C)$, $\mathcal{A}(\mathcal{S}) = \pi r^2$, if the triangle includes the whole circle [Fig. A.3(*a*)], (*b*) zero, $\mathcal{A}(\mathcal{S}) = 0$, otherwise, the triangle being outside the circle [Fig. A.3(*b*)];
- **Case 2** - If all the vertexes are outside the circle ($v_j \notin C$; $j = 1, ..., 3$) and one edge intersects the circle of $C$ in two points $a_1$ and $a_2$ [Fig. A.3(*c*)], the intersection region $\mathcal{S}$ is a circular surface delimited by the chord $\overline{a_1 a_2}$ and by the circular arc with endpoints $a_1$ and $a_2$ [Fig. A.3(*c*) - yellow area]. In this case, the value of $\mathcal{A}(\mathcal{S})$ is given by

$$\mathcal{A}(\mathcal{S}) = r^2 \arctan\left(\frac{\overline{a_1 a_2}}{2\sqrt{r^2 - \left(\frac{\overline{a_1 a_2}}{2}\right)^2}}\right) - \frac{\overline{a_1 a_2} \times \sqrt{r^2 - \left(\frac{\overline{a_1 a_2}}{2}\right)^2}}{2}. \quad (21)$$

Differently, if two or all edges intersect the circle [Figs. A.3(*d*)-A.3(*e*)], $\mathcal{S}$ is obtained as the difference between the circle area $\mathcal{A}(C)$ and the areas $\{\mathcal{A}(\mathcal{S}_i); i = 1, ..., I\}$ of the circular surfaces $\{\mathcal{S}_i; i = 1, ..., I\}$, which are computed as in (21),

$$\mathcal{A}(\mathcal{S}) = \mathcal{A}(C) - \sum_{i=1}^{I} \mathcal{A}(\mathcal{S}_i) \quad (22)$$

where $I$ indicates the number of circular surfaces and it is equal to $I = 2$ [Fig. A.3(*d*)] or $I = 3$ [Fig. A.3(*e*)] if the circumference is intersected by two or three edges, respectively;

- **Case 3** - If either one or two vertexes are inside the circle [Figs. A.3(*f*)-A.3(*h*)], $\mathcal{A}(\mathcal{S})$ is yielded as the sum of the area of the circular surface $\mathcal{S}_1$ [Fig. A.3(*f*)-A.3(*h*)], which is computed through (21), and that bounded by the polygon $\mathcal{F}$ whose vertexes are the $M$ intersection points $\{a_m; m = 1, ..., M\}$ between the circle $C$ and the edges of the triangle $\Delta$. The area of $\mathcal{F}$ is calculated by means of the the Gauss's formula [42] as follows

$$\mathcal{A}(\mathcal{F}) = \frac{1}{2} \left| \sum_{m=1}^{M-1} \Re(a_m) \Im(a_{m+1}) + \Re(a_M) \Im(a_1) - \sum_{m=1}^{M-1} \Re(a_{m+1}) \Im(a_m) - \Re(a_1) \Im(a_M) \right| \quad (23)$$

where $\Re(\cdot)$ and $\Im(\cdot)$ stand for the real and the imaginary parts, respectively;

- **Case 4** - If all the three vertexes lie within the circle, then the intersection surface $\mathcal{S}$ turns out out be equal to the triangle $\Delta$ and $\mathcal{A}(\mathcal{S}) = \mathcal{A}(\Delta)$ [Fig. A.3(*i*)] being

$$\mathcal{A}(\Delta) = \sqrt{\chi(\chi - e_1)(\chi - e_2)(\chi - e_3)} \quad (24)$$

where $\chi = \frac{e_1 + e_2 + e_3}{2}$ is the half of the length of triangle perimeter.

The procedure is iterated for all $T$ triangles against the $K + 1$ circles to compute the areas of all $K$ planar sectors, $\{\mathcal{A}\{[AF_k(u)]\}; k = 1, ..., K\}$ as well as the whole surface of $[AF(u)]$ such

$$\mathcal{A}\{[AF(u)]\} = \Sigma_{k=1}^{K} \mathcal{A}\{[AF_k(u)]\} \quad (25)$$

to be used in (12).

ACKNOWLEDGEMENTS

A. Massa wishes to thank E. Vico for her never-ending inspiration, support, guidance, and help.